\documentclass{aa}  
\usepackage{graphicx}
\usepackage{verbatim}
\usepackage{subcaption}
\usepackage{scalefnt}
\usepackage{txfonts}
\usepackage{xcolor}

\usepackage{placeins}

\newcommand{\rstar}{\ensuremath{r_{\mathrm{star}}}} 
\newcommand{\Mprimary}{\ensuremath{M_{\mathrm{p}}}}
\newcommand{\asecondary}{\ensuremath{a_{\mathrm{s}}}}
\newcommand{\esecondary}{\ensuremath{e_{\mathrm{s}}}}
\newcommand{\msecondary}{\ensuremath{m_{\mathrm{s}}}}
\newcommand{\rsecondary}{\ensuremath{r_{\mathrm{s}}}}

\begin{document} 

   \title{The resilience of the sailboat stable region}

   \author{Rafael Sfair\inst{1,2,3},
          Tiago F. L. L. Pinheiro\inst{2,4},
            Giovana Ramon\inst{2} \and
          Ernesto Vieira\inst{2}        
          }

    \institute{
    LIRA, Observatoire de Paris, Université PSL, Sorbonne Université, Université Paris Cité, CY Cergy Paris Université, CNRS,  92190 Meudon, France  \email{rafael.sfair@unesp.br} 
    \and
    São Paulo State University (UNESP), School of Engineering and Sciences, Guaratinguetá, SP, 12516-410, Brazil
    \and 
    Eberhard Karls Universität Tübingen,             Auf der Morgenstelle, 10, 72076, Tübingen, German
    \and
    Observatório Nacional (ON)/MCTI, Rio de Janeiro, RJ, 20921-400,Brazil}

      \date{Received \today; accepted ---}

\abstract
   {Binary systems host complex orbital dynamics where test particles 
   can occupy stable regions despite strong gravitational perturbations. 
   The sailboat region, discovered in the Pluto-Charon system, allows 
   highly eccentric S-type orbits at intermediate distances between the 
   two massive bodies. This region challenges traditional stability 
   concepts by supporting eccentricities up to 0.9 in a zone typically 
   dominated by chaotic motion.}
   {We investigate the sailboat region's existence and extent across 
   different binary system configurations. We examine how variations 
   in mass ratio, secondary body eccentricity, particle inclination, 
   and argument of pericenter affect this stable region.}
   {We performed 1.2 million numerical simulations of the elliptic 
   three-body problem to generate four datasets exploring different 
   parameter spaces. We trained XGBoost machine learning models to 
   classify stability across approximately $10^9$ initial conditions. 
   {We validated our results using Poincaré surface of section and Lyapunov 
   exponent analysis to confirm the dynamical mechanisms underlying the stability}.}
   {The sailboat region exists only for binary mass ratios 
   $\mu = [0.05, 0.22]$. Secondary body eccentricity severely constrains 
   the region, following an exponential decay: 
   $e_{s,\mathrm{max}} \approx 0.016 + 0.614 \exp(-25.6\mu)$. 
   The region tolerates particle inclinations up to $90^\circ$ and 
   persists in retrograde configurations for $\mu \leq 0.16$. 
   Stability requires specific argument of pericenter values within 
   $\pm 10^\circ$ to $\pm 30^\circ$ of $\omega = 0^\circ$ and $180^\circ$. 
   Our machine learning models achieved over 97\% accuracy in predicting 
   stability.}
   {The sailboat region shows strong sensitivity to system parameters, 
   particularly secondary body eccentricity. Among Solar System dwarf 
   planet binaries, Pluto-Charon, Orcus-Vanth and Varda-Ilmarë systems could harbor 
   such regions. The combination of numerical simulations and machine 
   learning provides an efficient approach for mapping stability in 
   complex gravitational systems.}


   \maketitle

\section{Introduction}
\label{S-Introduction}

Binary systems,
where two massive bodies orbit their common center of mass, represent fundamental configurations in celestial mechanics. 
Understanding the dynamics of test particles within these systems has direct applications
for planetary formation, spacecraft mission design, satellite detection,
and the study of natural debris distributions.

Among the various orbital configurations possible in binary
systems the S-type orbits, where particles orbit one of the massive
bodies, exhibit complex stability patterns. These patterns depend on
multiple parameters including mass ratio, eccentricity, and orbital
inclination.

The Pluto-Charon system provides an exceptional natural laboratory for
studying binary dynamics. With a mass ratio of $\mu \sim 0.12$ and a separation
distance of $d = 19,595$~km \citep{Brozovic2024}, this system challenges
traditional models due to Charon's significant mass relative to Pluto.
The system also hosts four small satellites (Styx, Nix, Kerberos, and Hydra)
in P-type orbits around the system's barycenter, demonstrating the
richness of its dynamical environment.

Previous investigations of the Pluto-Charon system revealed unexpected
stable regions for test particles. \citet{Winter2010, Winter2013,
Giuliatti2014} made extensive numerical simulations to map S-type
orbital stability around both Pluto and Charon. Their works identified
three distinct stable regions for prograde particles: the first around Charon
extending to $0.2d$ with eccentricities below 0.6, the second region lies close to Pluto and extends to $0.5d$, while the
third region named ``sailboat region'', due to its distinctive shape
in $(a,e)$ parameter space, occupies the mid-distance between Pluto and
Charon with high eccentricities ranging from 0.2 to 0.9.

Unlike typical stable
regions that favor near-circular orbits or occupy mean-motion resonances,
this region allows highly eccentric orbits at intermediate distances.
\citet{Giuliatti2014} demonstrated through Poincaré surfaces of section that
the sailboat region corresponds to a family of periodic orbits (family
``BD'') derived from the circular restricted three-body problem
\citep{Brouke1968}. Further analysis showed that the region persists for
inclinations up to $90^\circ$ and exists within specific intervals of the
argument of pericenter: $\omega = \left[-10^\circ, 10^\circ\right]$
and $\left[160^\circ, 200^\circ\right]$, reaching
maximum extent at $\omega = 0^\circ$ and $180^\circ$.

This work investigates the resilience of the sailboat region by
exploring how it responds to variations in key system parameters.
We examine the effects of changing the binary mass ratio, secondary
body eccentricity, particle inclination, and argument of pericenter
on the existence and extent of this stable region. To map the vast
parameter space efficiently, we combine direct numerical simulations
with machine learning algorithms, {so our work focuses on numerically 
mapping the sailboat region's extent across multiple parameter spaces 
rather than pursuing analytical solutions, which remain challenging for 
such complex multi-parameter dynamical systems.} 
A similar study was applied by \citet{Pinheiro2025}, 
who performed machine learning techniques to map the stability of a star-planet-particle test system, 
achieving an accuracy of 98.48$\%$.
We first generate training datasets
through targeted three-body simulations, then apply supervised learning
to predict stability across billions of initial conditions that would
be computationally prohibitive to simulate directly.

{The paper is organized as follows. Section~\ref{S-Binaries} examines different types
of binary systems and their characteristics, and Sec.~\ref{S-Methods} describes our
numerical simulation methods and machine learning approach. Section~\ref{S-Massratio}
analyzes how the sailboat region varies with binary mass ratio, while
Section~\ref{S-DifferentInitialconditions} investigates the effects of secondary body eccentricity,
particle inclination, and argument of pericenter. Section~\ref{S-Poincare} validates our results using 
stability verification methods including Poincaré surface of section and Lyapunov exponent analysis.
The last section presents our remarks.}

\section{Binary systems}
\label{S-Binaries}

To understand the broader context of sailboat regions beyond the 
Pluto-Charon system, we examine binary configurations across different 
astrophysical environments. This analysis helps identify which types of 
binary systems could potentially harbor sailboat regions and provides 
observational targets for future studies. Since the sailboat region 
requires specific mass ratio ranges and orbital configurations, 
understanding the distribution of binary system properties is important  
for predicting where such regions might exist.

Binary star systems can be classified into three distinct types based on 
the separation distance between the components and their physical sizes.
Detached systems have both stars well within their Roche lobes, maintaining 
stable orbital configurations without mass exchange. Semi-detached systems 
have one star that has filled its Roche lobe, enabling mass transfer 
to its companion and creating dynamically active environments. Contact 
systems have both stars filling their Roche lobes, resulting in continuous 
mass exchange between the components \citep{Kopal1955}.

\begin{figure}[h]
    \centering
    \includegraphics[width=0.45\textwidth]{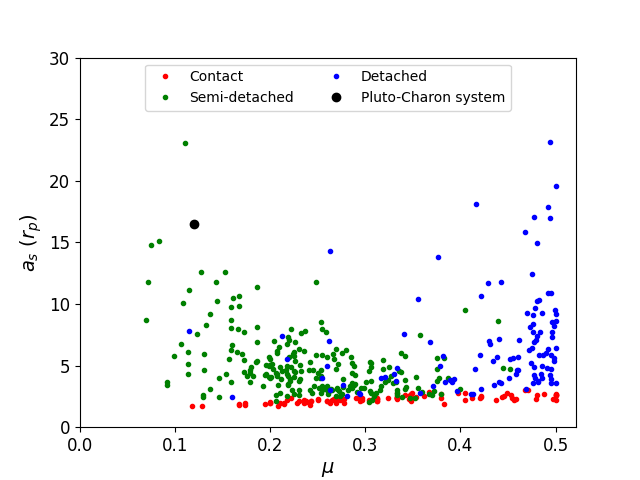}
    \caption{Diagram of a binary star system ($d \times \mu$), 
    where $d$ represents the separation distance between the stars
    and $\mu$ is the binary mass ratio. The diagram categorizes 
    the binaries as follows: blue for Detached type, 
    green for Semi-Detached type, and red for Contact type. 
    The black point corresponds to Pluto-Charon system.}
    \label{F-Stars}
\end{figure}

Figure~\ref{F-Stars} presents the distribution of semi-major axis 
(\asecondary) in units of the primary star radius (\rstar) versus 
binary mass ratio ($\mu$) for 489 eclipsing binary systems from 
observational surveys. The sample has 100 contact systems, 
232 semi-detached systems, and 157 detached systems, compiled from 
three catalogs: the catalog of DMS-type eclipsing binaries 
\citep{Svechnikov2004vizier}, catalog of early-type contact binaries 
\citep{Bondarenko2004vizier}, and semi-detached eclipsing binaries 
\citep{Surkova2004vizier}.

The majority of binary systems with mass ratios comparable to 
Pluto-Charon (indicated in black) are semi-detached configurations, 
as shown in Figure~\ref{F-Stars}. The ongoing mass transfer processes 
in these systems create unstable dynamical environments in the 
circumbinary region. However, four detached systems exhibit $\mu < 0.25$ 
with separation distances ranging from 2 to 8 \rstar. 
{These detached configurations provide stable dynamical 
environments that can harbor sailboat regions, consistent 
with the findings of \cite{Giuliatti2014} . 
Similar stability considerations apply to smaller 
binary systems within our Solar System, particularly among 
Trans-Neptunian dwarf planet binaries.}

The term ``dwarf planet'' was introduced during the XXVIth General 
Assembly of the IAU. Unlike planets, this new category of objects 
has not cleared the surrounding region of their orbits. 
Table \ref{T-Dwarf} shows the ten largest confirmed or candidate 
dwarf planets that host at least one satellite. 
This table includes the mass ratio of the system, 
orbital semi-major axis (\asecondary) and eccentricity (\esecondary) of each satellite.

{All these bodies are Trans-Neptunian objects (TNOs). }
Among the confirmed dwarf planets, only Ceres and Sedna 
have no discovered satellites, while Pluto hosts five satellites, 
and Haumea hosts two: Hi'iaka and Namaka.

\begin{table}[ht]
\scalefont{0.85}
\caption{ Binary systems of dwarf planet and their satellites in our Solar System.}
\begin{tabular}{ccccc}
\hline
{Binary System}                 & {$\mu$}        & {\asecondary}  (km) & {\esecondary}       \\ \hline 
Pluto and Charon$^{a}$                 & 0.12                  & 19600               & 0.000161           \\ 
Eris and Dysnomia$^{b,c}$              & 0.0050                & 37273               & 0.0062             \\
Haumea and Hi'iaka$^{d}$               & 0.012                 & 49880               & 0.0513             \\ 
Makemake and S/2015 (136472) 1$^{d}$   & 0.002                 & 21100               & unknown            \\ 
Quaoar and Weywot$^{e}$                & 0.00433               & 13289               & 0.056              \\ 
Gonggong and Xiangliu$^{f}$            & unknown               & 24021               & 0.2908             \\ 
Orcus and Vanth$^{b}$                  & 0.16                  & 8999.8              & 0.00091            \\ 
Salacia and Actaea$^{g}$               & 0.033                 & 5724                & 0.0098             \\ 
Varda and Ilmarë$^{h}$                 & 0.083                 & 4809                & 0.0181             \\ 
(532037) 2013 FY27 $^{i}$        & unknown               & 9800                & unknown            \\ \hline        
\end{tabular}
\tablebib{
        (a) \cite{Brozovic2024}, 
        (b) \cite{Brown2023}, (c) \cite{Holler2021}, 
        (d) \cite{Ragozzine2009}, (e) \cite{Morgado2023}, 
        (f) \cite{Kiss2019}, (g) \cite{Grundy2019}, 
        (h) \cite{Grundy2015}, (i) \cite{Sheppard2018}.  
        {Unknown values indicated where observational data are unavailable.}}
\label{T-Dwarf} 
\end{table}

Moreover, Quaoar and Haumea have ring systems that orbit 
close to the 1:3 spin–orbit resonance with their primary body. 
In the case of Quaoar, its two rings are located outside its Roche limit
\citep{Pereira2023, Morgado2023, Ortiz2017}.

The Pluto-Charon system is the only Trans-Neptunian dwarf planet 
that has been visited by a spacecraft. 
Two systems, Orcus-Vanth and Varda-Ilmarë, have $\mu$ values 
close to that of Pluto-Charon, with $\mu = 0.16$ and 0.083, 
respectively. 
This suggests that both binaries likely exhibit a sailboat 
region similar to the Pluto system.
Investigating stability regions around these bodies can be 
helpful for future space missions and observation campaigns
to identify the most probable locations to discover new 
objects or ring systems.

{Having identified potential binary systems that could 
harbor sailboat regions, we now examine the computational methods used to map their stability.}

\section{Methods}
\label{S-Methods}

\qquad
To investigate the stability of the sailboat region across different 
binary system configurations and orbital parameters, we implemented two 
complementary approaches. First, we performed direct numerical 
simulations of the three-body problem as described in 
Sec.~\ref{S-NumericalSimulations}. 
These simulations generated both stability maps and training data for our 
machine learning models. Then, we applied supervised learning algorithms 
to classify stability across a broader parameter space, as detailed in 
Section~\ref{S-MachineLearning}. This combined methodology allowed us to 
efficiently explore stability conditions that would be computationally 
prohibitive through direct simulation alone.

\subsection{Numerical method}
\label{S-NumericalSimulations}

\qquad
To characterize the sailboat region across different parameter spaces, we 
performed extensive numerical simulations of the elliptic three-body problem. 
We modeled a dimensionless binary system with test particles in S-type orbits 
around the primary body, subject to gravitational perturbations from the 
secondary body.

In our simulations, we defined the primary body as a point mass with 
$\Mprimary = 1 - \mu$, where $\mu$ represents the binary mass ratio. The 
secondary body has mass $\msecondary = \mu$, semi-major axis $\asecondary = 1$, 
and collision radius ($\rsecondary$) defined as:

\begin{align}
    r_{s} = 0.1 \left[ \left(1 - e_\mathrm{s} \right)\sqrt[3]{\frac{\mu}{3}} \right]
    \label{E-radius}
\end{align}

\noindent where $\esecondary$ is the secondary body's eccentricity, 
initially set to zero but varied in later simulations as described in 
Sec~\ref{Ss-eccentricity}. This collision radius formulation accounts for 
the Hill sphere of the secondary body, scaled by a factor of 0.1 to 
represent the effective collision cross-section. The $(1 - e_\mathrm{s})$ 
term adjusts for orbital eccentricity, ensuring that the collision radius 
remains appropriate for eccentric secondary orbits. The $\sqrt[3]{\mu/3}$ 
factor represents the Hill radius normalization in the three-body problem.
{This formulation ensures that particles approaching within the secondary's 
gravitational sphere of influence are considered collided, preventing artificial close encounters.}

We initially simulated 300,000 binary systems, randomly selecting the mass ratio 
$\mu$ from a uniform distribution in the range $[0.01 - 0.3]$. For test particles 
in these systems, we also randomly selected orbital parameters from uniform 
distributions: semi-major axis $a = [0.45 - 0.7]$ and eccentricity $e = [0 - 0.99]$. 
All other orbital elements of both the secondary body and test particles were 
set to zero. Using the \textsc{Rebound} package with the IAS15 integrator 
\citep{Hanno14}, we integrated each system for $10^4 \times 2\pi$ time units, 
following \citet{Winter2001} who demonstrated this duration is sufficient to 
establish orbital stability.

Simulations terminated when particles either collided with the secondary body 
or were ejected from the system (defined as $e \geq 1$ or $a > 1$). For each 
initial condition, we recorded the outcome (stable or unstable) 
and the termination cause when applicable. 
This complete dataset, containing both stable and unstable particles 
with their respective parameters, formed the 
training data for our machine learning algorithms.

To explore additional parameter dependencies, we generated three more datasets 
of 300,000 simulations each, systematically varying the secondary body eccentricity 
($\esecondary$), particle inclination ($i$), and argument of pericenter 
($\omega$). Table \ref{T-Dataset} summarizes these datasets and their 
objectives.

\begin{table}[ht]
  \scalefont{0.9}
    \caption{Description of dataset parameters.}

  \begin{center}
  \begin{tabular}{c c c}
  \hline
  {Dataset} & {Features} & {Parameter investigated} \\ \hline
  A & $\mu$, a, e & Binary mass ratio \\ \hline
  B & $\mu$, a, e, $\esecondary$ & Secondary body eccentricity \\ \hline
  C & $\mu$, a, e, i & Particle inclination \\ \hline
  D & $\mu$, a, e, $\omega$ & Particle argument of pericenter \\ \hline
  \end{tabular}
  \label{T-Dataset}
  \end{center}
\end{table}

For each dataset, we optimized machine learning models through cross-validation, 
class imbalance correction, and parameter tuning. The resulting models allowed
us to predict stability for more than 3.6 billion initial conditions, far beyond what
would be computationally feasible through direct simulation.

\subsection{Machine Learning approach}
\label{S-MachineLearning}

\qquad
We employed the XGBoost algorithm (Extreme Gradient Boosting) to classify
orbital stability. This supervised learning method performs binary 
classification, categorizing particles as either stable or unstable based 
on their orbital parameters.
The effectiveness of this algorithm in classifying stable orbits has 
been demonstrated in previous studies, such as \cite{Tamayo2016} and \cite{Pinheiro2025}.

XGBoost builds an ensemble of decision trees through sequential training, 
where each new tree corrects errors from previous iterations 
\citep{Geron2022hands}. The algorithm computes residuals for the predictions
and refines the model through successive iterations \citep{Wade2020hands}, 
making it particularly effective for our classification task.

We divided our data set into training and validation (80\% of the data), 
and test (20\%) sets. The training data established predictive patterns, the
validation data optimized the model parameters, and the test data provided an
unbiased performance evaluation \citep{Shalev2014}. We implemented 
k-fold cross-validation by partitioning the training data into subsets, 
using each subset iteratively for validation while training on the 
remaining data. This technique prevents overfitting and improves 
generalization \citep{Hastie2009}.

Model optimization involved tuning both hyperparameters and classification 
thresholds. We selected optimal hyperparameters through grid search 
\citep{Weerts2020}. For classification thresholds, we adjusted the
default 0.5 probability cut-off to address class imbalance in our datasets, 
as stability cases were significantly outnumbered by instability cases
in certain parameter regimes.

\section{Analysis of Sailboat region for different binary mass ratios}
\label{S-Massratio}

The first dataset consisted of three freedom parameters: $\mu$, $a$ and $e$. 
This stage aimed to verify how the sailboat region changes with varying 
mass ratios of the binary system.

This dataset contained 300,000 samples with the following distribution: 
77.98\% resulted in collisions, 0.75\% in particle ejection, and 21.27\% 
remained stable throughout the simulations. To address this imbalanced 
class distribution, we implemented a random oversampling technique, 
selecting samples from the minority class and adding them to the training 
dataset.

Our best model achieved 98.53\% accuracy, measuring the ratio of correctly 
predicted outcomes. For imbalanced classification problems, we evaluated 
additional performance metrics: recall and precision. Recall measures the 
ratio of correctly identified samples within each class, reaching 97\% for 
stable and 99\% for unstable classes. Precision measures the ratio of 
correct classifications within each predicted class, achieving 97\% for 
stable and 99\% for unstable classes. 
{This accuracy significantly exceeds the $\sim90$\% threshold 
typically considered acceptable for astronomical classification tasks.}

\begin{figure*}[!ht]
    \centering
    \includegraphics[width=0.7\textwidth]{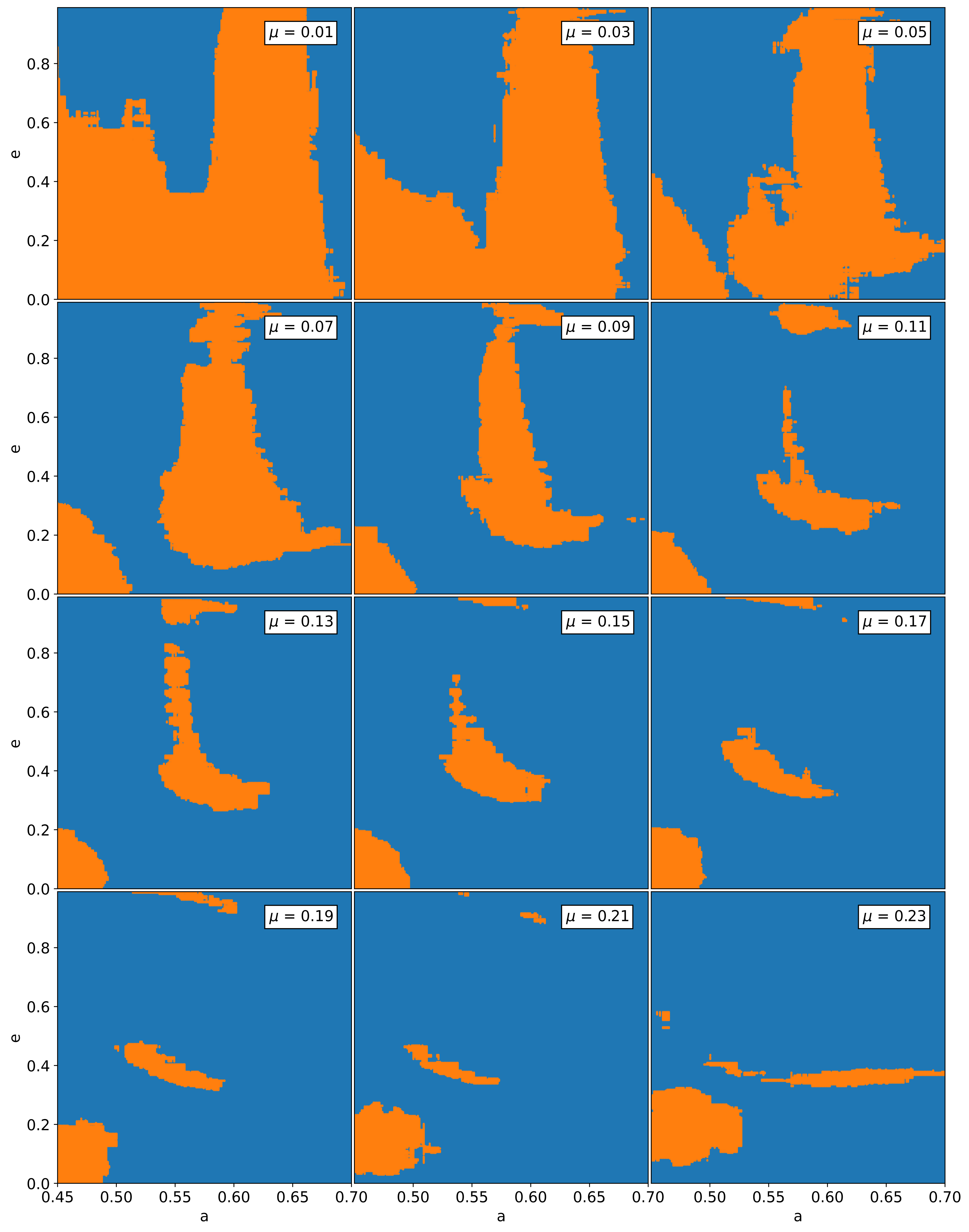}
    \caption{Stability maps in the $(a,e)$ plane for binary systems with mass ratios 
    $\mu = [0.01-0.23]$. Stable initial conditions are shown in orange, unstable in blue. 
    All particles are in coplanar orbits around the primary body with 
$\omega = \Omega = f = 0^\circ$.}
\label{F-Massratio}
\end{figure*}

Figure~\ref{F-Massratio} presents stability maps derived from machine 
learning predictions for binary systems with mass ratios in the range
$\mu = [0.01 - 0.23]$. Each diagram represents 247,500 different initial 
conditions, with particle semi-major axis $a = [0.45 - 0.7]$ and eccentricity 
$e = [0 - 0.99]$ using step sizes of $\Delta a = \Delta e = 0.001$. Our 
complete analysis covered the range $\mu = [0.01 - 0.30]$, revealing that 
the sailboat region disappears entirely for $\mu \geq 0.24$, placing 
an upper limit on the mass ratio for the existence of this region.

\FloatBarrier

As shown in Figure~\ref{F-Massratio}, binary systems with $\mu \leq 0.04$ have a 
contiguous stable region around the primary body extending up to $\sim 0.7$. 
In the interval $a = [0.55 - 0.68]$, stable particles can reach high 
eccentricities, though they remain below $e = 1$ due to the stability 
criterion.

For $\mu = 0.05$, the stability map reveals two distinct regions:
(i) one confined to $a < 0.52$ and $e \leq 0.4$, and 
(ii) the sailboat region located at $a = [0.52 - 0.7]$.
This represents the smallest mass ratio where we can identify 
two separated stable regions, marking where the sailboat region 
detaches from the inner stability region. As $\mu$ increases, the 
sailboat region progressively diminishes in size until it completely 
vanishes at $\mu = 0.22$.

The minimum particle eccentricity of the sailboat region 
increases with higher $\mu$ values. For a binary system with $\mu = 0.07$, 
sailboat particles have eccentricity $e > 0.1$, while for $\mu = 0.09$ 
the eccentricities increase to $e > 0.18$. When $\mu = 0.10$, the 
sailboat splits into two regions: the larger one at 
$a = [0.52 - 0.67]$ and $e = [0.2 - 0.8]$, while the smaller region 
contains particles with highly eccentric orbits ($e > 0.85$).

\begin{figure}[h]
    \centering
    \includegraphics[width=0.45\textwidth]{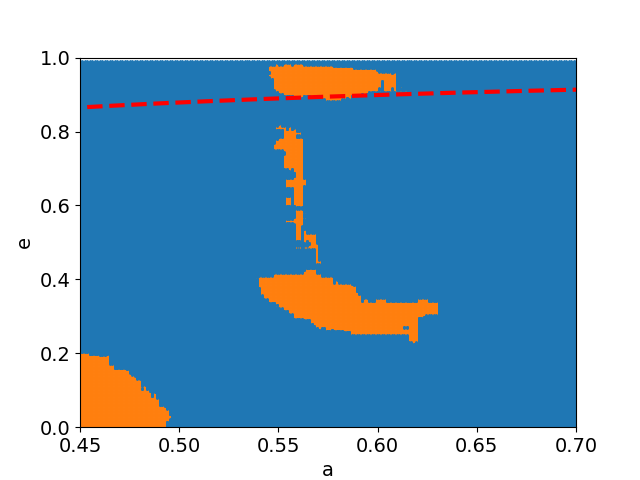}
    \caption{The sailboat region for a binary system with $\mu = 0.12$
    (the closest case with Pluto-Charon system, where Charon has eccentricity).
    The red dashed line represents the pericenter of 
    the particle in the same distance of Pluto's radius.
    There is other stable regions in the range $a = [0.45 - 0.7]$,
    not showed in the plot.}
    \label{F-Pluto}
\end{figure}

Some eccentric stable orbits ($e > 0.85$) in the sailboat region can 
become unstable depending on the size of the primary body. 
Figure~\ref{F-Pluto} shows the sailboat region for a binary with $\mu = 0.12$ 
(similar to the Pluto-Charon system). The red dashed line indicates when 
the pericenter of the particle has the same distance as Pluto's radius. 
As shown in Figure~\ref{F-Pluto}, all particles with $e > 0.9$ 
(above the red dashed line) will collide with Pluto, which matches 
the results of \citet{Giuliatti2014}.

For binary systems with $\mu = 0.15$, the main region in the sailboat 
is constrained to a range of $a = [0.52 - 0.63]$ and 
$e = [0.22 - 0.7]$, becoming less eccentric and losing its sailboat-like 
shape. When $\mu = 0.16$ the maximum eccentricity for this 
region is $\sim 0.6$, and for $\mu = 0.18$, it is situated at 
$a = [0.5 - 0.6]$ with $e < 0.5$, showing a gradual shift towards
the primary body and a decrease in the maximum eccentricity.

For $\mu = 0.22$, a new stable region emerges at $a > 0.65$ and 
$e \sim 0.4$. This region expands in size and merges with the 
remaining sailboat at $\mu = 0.23$.

\section{Different initial conditions}
\label{S-DifferentInitialconditions}

\subsection{Eccentric secondary body case}
\label{Ss-eccentricity}

We investigate how the secondary body's eccentricity affects 
the sailboat region stability. This dataset contains four parameters: 
binary mass ratio ($\mu = [0.01 - 0.23]$), particle semi-major axis 
($a = [0.45 - 0.7]$), particle eccentricity ($e = [0 - 0.99]$), and 
secondary body eccentricity ($e_s = [0 - 0.99]$).

The overall outcome from 300,000 simulated samples shows that 1.77\% 
of the particles remain stable, while 98.23\% become unstable. Among 
the unstable particles, 95.39\% are ejected and 2.84\% collide with 
the secondary body. Our best machine learning model utilizes the 
ADASYN (Adaptive Synthetic Sampling) oversampling technique to handle 
this significant class imbalance, where the ratio between minority and 
majority classes is approximately 1:50.

The ADASYN method generates synthetic samples for the minority class 
in the training dataset based on the distribution of samples belonging 
to that class in the feature space \citep{He2008}. Our best machine 
learning model achieves an accuracy of 99.55\%. Specifically, for the 
stable class, we obtain a recall and precision of 87\%, and almost 
100\% for the unstable class. From 60,000 samples in the test dataset, 
the model correctly classifies 913 as stable and 58,817 as unstable, 
misclassifying only 270 particles (135 from the stable class 
classified as unstable, and 135 from the unstable class classified 
as stable).

\begin{figure}[h]
    \centering
    \includegraphics[width=0.45\textwidth]{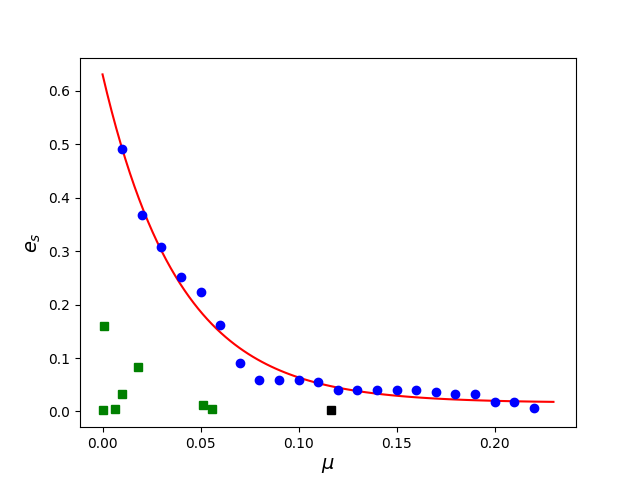}
    \caption{Maximum eccentricity of secondary body (blue dots)
    versus mass ratio of the system. 
    The red curve represents a fitted model obtained using the 
    Least Squares method.
    Green squares denote binary systems of dwarf planets 
    and their satellites, while the black square represents 
    the Pluto-Charon system.}
    \label{F-Ecc}
\end{figure}

We employ our machine learning model to predict stability maps 
$(a \times e)$ for each $\mu \leq 0.22$ and determine the maximum 
secondary body eccentricity value where the sailboat region 
completely vanishes, using a step size of $\Delta e_s = 0.001$. 
Figure~\ref{F-Ecc} shows the maximum eccentricity of the secondary 
body where the sailboat region exists (blue dots) for different mass 
ratios. The red curve represents a fit obtained through the least 
squares method. Each green square represents a binary system of dwarf 
planets and their satellites shown in Table~\ref{T-Dwarf}, while the 
black square indicates the Pluto-Charon system.

\FloatBarrier

Small changes in the secondary body eccentricity are sufficient to 
substantially decrease the extension of the sailboat region. An 
empirical expression for the maximum eccentricity value 
($e_{s,\mathrm{max}}$) as a function of $\mu$ can be expressed as

\begin{align}
    e_{s,\mathrm{max}} \approx (0.016 \pm 0.006) + 
    (0.614 \pm 0.023)e^{(-25.626 \pm 1.636)\mu}
    \label{E-ecc}
\end{align}

\noindent This equation indicates a strong exponential decay between 
$e_{s,\mathrm{max}}$ and the $\mu$ parameter. Figure~\ref{F-Ecc} shows 
that all dwarf planets present in Table~\ref{T-Dwarf} may exhibit a 
sailboat region, indicating potential regions for hosting satellites 
or ring systems. It is even possible to design missions with 
spacecraft studying the system from an orbit within this region due 
to its stability.

\begin{center}
\begin{figure*}[!ht]
    \begin{subfigure}{0.88\linewidth}
        \centering
        \includegraphics[width=0.8\linewidth]{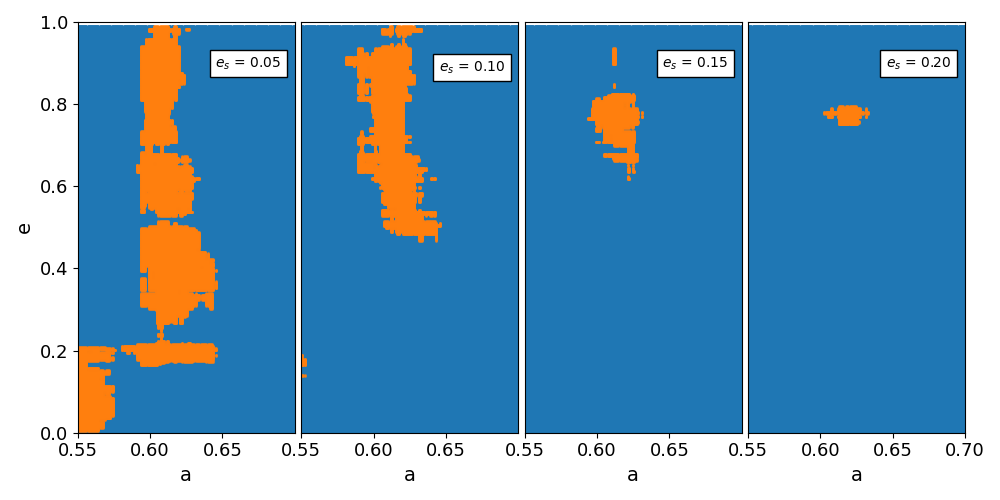}
        \caption{$\mu = 0.05$}
    \end{subfigure}
    \begin{subfigure}{0.88\linewidth}
        \centering
        \includegraphics[width=0.8\linewidth]{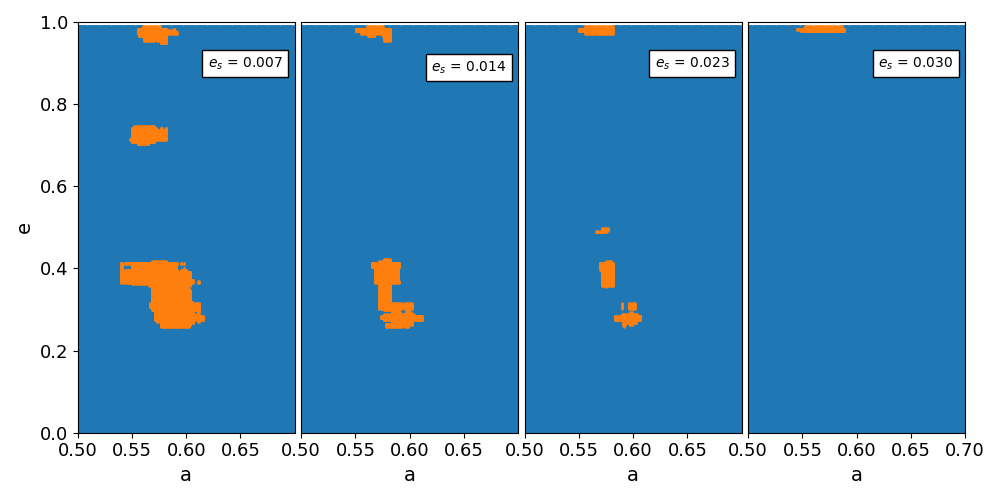}
        \caption{$\mu = 0.12$}
    \end{subfigure}
     \begin{subfigure}{0.88\linewidth}
        \centering
        \includegraphics[width=0.8\linewidth]{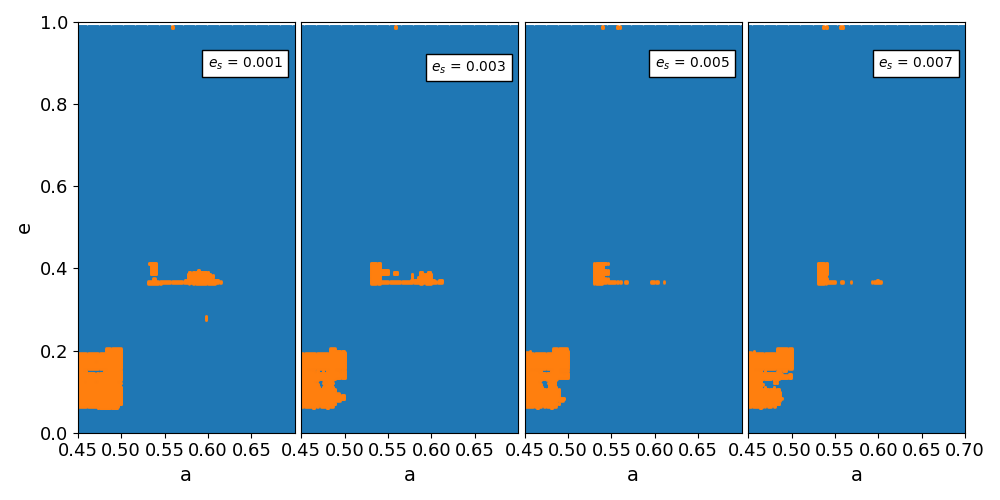}
        \caption{$\mu = 0.22$}
    \end{subfigure}
    \caption{Stability maps $(a \times e)$ showing the 
evolution of the sailboat region with changing secondary 
body eccentricity for binary systems with mass ratios 
of $\mu = 0.05$, $0.12$, and $0.22$.}
    \label{F-Seqecc}
\end{figure*}
\end{center}

Figure~\ref{F-Seqecc} shows the evolution of the sailboat region as 
the secondary body eccentricity changes for binary systems with mass 
ratios of $\mu = 0.05$, $0.12$, and $0.22$.

Particles with lower eccentricity within the sailboat region become 
unstable as the eccentricity of the secondary body increases. In 
Figure~\ref{F-Seqecc}, for $\mu = 0.05$ and $e_s = 0.2$, only 
particles with $e \sim 0.8$ survive. When $\mu = 0.12$ with 
$e_s = 0.03$ and $\mu = 0.22$ with $e_s = 0.007$, only particles 
with $e$ approaching unity survive, however they will probably 
collide with the primary body.

When disregarding particles with $e > 0.9$ to prevent collisions 
with the primary body, the maximum limit of secondary body 
eccentricity shows significant changes for $\mu \geq 0.12$ compared 
with the results shown in Figure~\ref{F-Ecc}. For $\mu$ values 
ranging from 0.12 to 0.16, the maximum eccentricity is 0.04. 
However, for $\mu = 0.17$, $e_{s,\mathrm{max}}$ decreases to 0.036, 
a difference of 0.004 when considering that stable particles
exceeding eccentricity 0.9 become unstable. For $\mu = 0.18$ and 
$0.19$, the value of $e_{s,\mathrm{max}} = 0.033$, while for 
$\mu = 0.20$ and $0.21$, $e_{s,\mathrm{max}}$ reaches 0.018.

{These results demonstrate that secondary body 
eccentricity imposes the strongest constraint on sailboat region 
existence, with tolerance decreasing exponentially as $\mu$ increases.}

\subsection{Inclined particle case}
\label{Ss-inclination}

\qquad
To analyze how particle inclination affects the sailboat region, 
we generated a third dataset of 300,000 samples with four parameters: 
binary mass ratio ($\mu = [0.01 - 0.23]$), particle semi-major axis 
($a = [0.45 - 0.7]$), particle eccentricity ($e = [0 - 0.99]$), and 
orbital inclination ($i = [0^\circ - 180^\circ]$).

This dataset shows a different class distribution compared to the 
eccentric secondary body case, with 20.11\% stable particles and 
79.89\% unstable particles. Among the unstable cases, 79.75\% result 
in ejection and only 0.14\% record collisions. The higher proportion 
of stable particles compared to the previous dataset reduces the 
class imbalance problem, though oversampling techniques remain 
necessary for optimal model performance.

Our best machine learning model employs SMOTE (Synthetic Minority 
Oversampling Technique) to handle the remaining class imbalance. 
SMOTE creates synthetic samples within the training data by 
generating new instances for the minority class based on the 
euclidean distance between nearest neighbors, using a multiplicative 
factor between 0 and 1 \citep{Chawla2002}. This approach differs 
from ADASYN by focusing on the feature space structure rather than 
the local density distribution.

\begin{figure}[!h]
    \centering
    \includegraphics[width=0.5\textwidth]{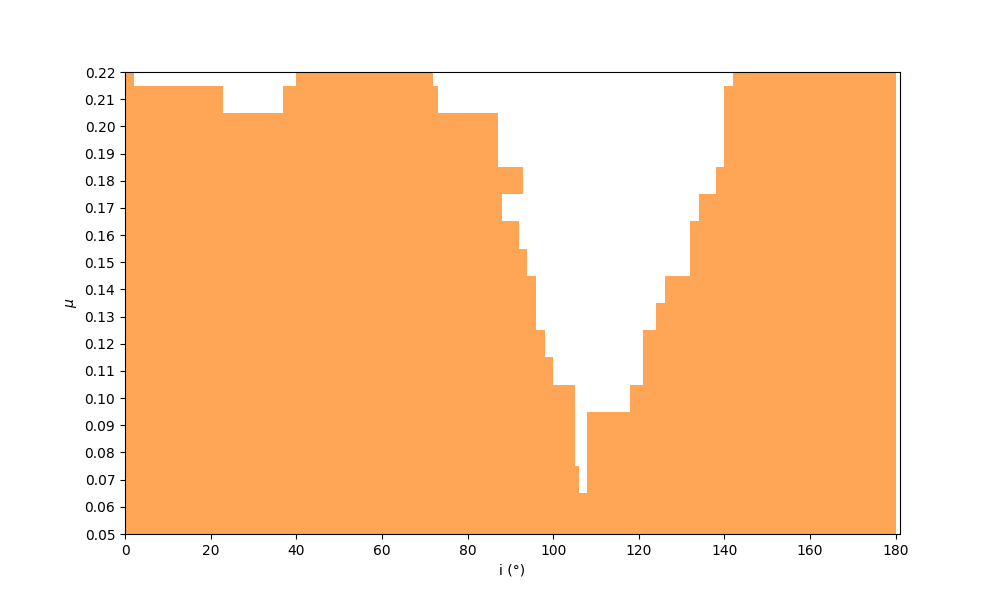}
    \caption{Stability regions for different particle orbital 
    inclinations. Orange areas indicate inclination values where 
    stable regions exist within $a = [0.4 - 0.7]$ for various $\mu$.
    {The gaps in coverage indicate parameter combinations where the sailboat region becomes unstable. } }
    \label{F-InclinationMap}
\end{figure}

The optimized model achieves an accuracy of 97.15\%, with precision 
and recall of 93\% for the stable class and 98\% for the unstable 
class. Using this model, we predict stability diagrams in the 
$(a \times e)$ space from $i = 0^\circ$ to $180^\circ$, with each 
step incrementing by $\Delta i = 1^\circ$. This analysis covers 
each $\mu \leq 0.22$ with step size $\Delta \mu = 0.01$. 
Figure~\ref{F-InclinationMap} shows where stable regions exist within 
the range $a = [0.45 - 0.7]$ for different values of $\mu$ and 
particle orbital inclination.

For the highest mass ratio, $\mu = 0.22$, even a $2^\circ$ inclination 
eliminates stability {within} $a = [0.45 - 0.70]$. At $\mu = 0.21$, the region 
persists up to $i \approx 20^\circ$, disappears, then reappears at 
$i \approx 35^\circ$ and extends to $i \approx 70^\circ$. The gap at 
$i < 35^\circ$ may indicate limited model generalization due to sparse 
stable particles in this parameter space. Retrograde regions appear at 
$i \approx 140^\circ$ and reach maximum extent at $i = 180^\circ$, 
though only for $\mu < 0.17$.

For mass ratios between 0.17 and 0.20, the sailboat region survives 
up to $i = [85^\circ - 95^\circ]$. When $\mu \leq 0.16$, the region 
remains stable for $i = [90^\circ - 100^\circ]$, consistent with 
\cite{Winter2013} findings for the Pluto-Charon system. For 
$\mu = [0.07 - 0.10]$, stability extends to $i \sim 105^\circ$, 
with retrograde regions appearing at progressively lower inclinations 
as $\mu$ decreases.

\begin{center}
    
\begin{figure*}[!ht]
    \begin{subfigure}{0.85\linewidth}
        \centering
        \includegraphics[width=0.8\linewidth]{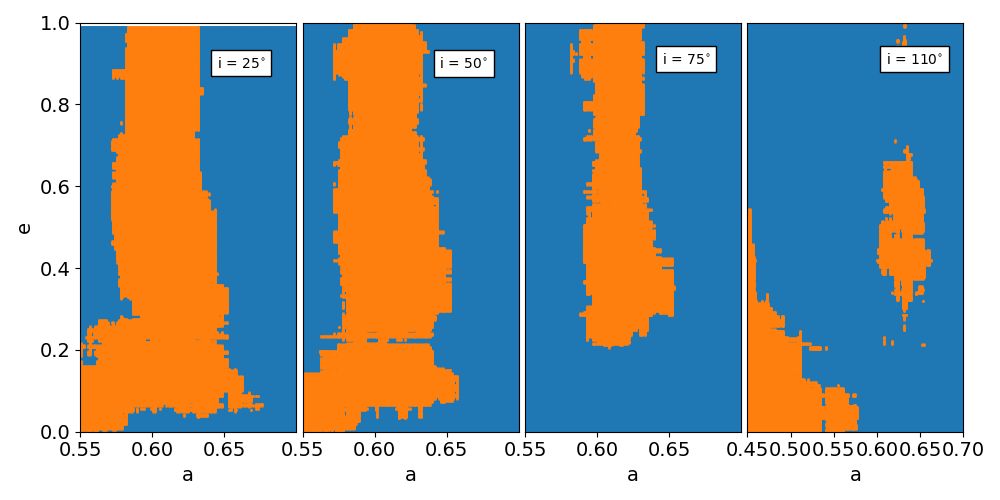}
        \caption{$\mu = 0.05$}
    \end{subfigure}
    \begin{subfigure}{0.85\linewidth}
        \centering
        \includegraphics[width=0.8\linewidth]{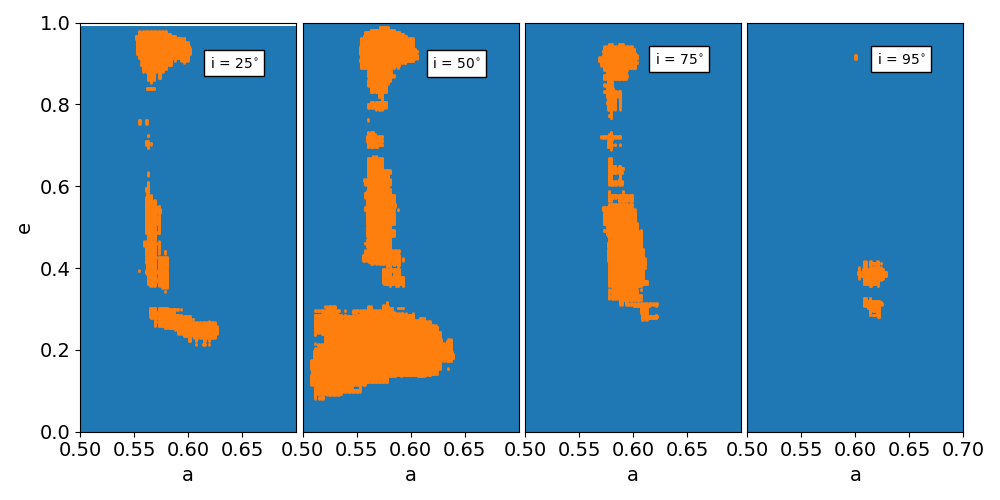}
        \caption{$\mu = 0.12$}
    \end{subfigure}
    \caption{Stability maps $(a \times e)$ showing the 
evolution of the sailboat region for varying particle inclinations 
in binary systems with mass ratios $\mu = 0.05$ and $0.12$.}
    \label{F-Seqinc}
\end{figure*}
\end{center}

For the smallest mass ratios, $\mu = 0.05$ and $0.06$, the sailboat 
region shows continuous existence across all inclinations, reaching 
minimum size at $i = 110^\circ$ before expanding again toward 
retrograde inclinations. Figure~\ref{F-Seqinc} illustrates the 
sailboat evolution for $\mu = 0.05$ and $0.12$ at representative 
inclinations. These results show that the strong resilience of the sailboat 
to inclination changes. 

\subsection{Varying the argument of pericenter}
\label{Ss-omega}

Our final dataset explores how the argument of pericenter affects 
sailboat region stability. We simulated 300,000 systems with four 
parameters: binary mass ratio ($\mu = [0.01 - 0.23]$), particle 
semi-major axis ($a = [0.45 - 0.7]$), particle eccentricity 
($e = [0 - 0.99]$), and argument of pericenter 
($\omega = [0^\circ - 360^\circ]$).

The simulation outcomes show 11.5\% of particles remaining stable throughout 
the integration period, while 87.3\% experienced ejection and 1.2\% 
collided with the secondary body. This class distribution presents a 
moderate imbalance compared to the eccentric secondary body case.

Our optimized machine learning model employed random oversampling to 
address class imbalance, achieving 98.5\% accuracy. The stable class 
achieved 94\% recall and 93\% precision, while the unstable class reached 
99\% for both metrics, demonstrating reliable classification performance 
across parameter space.

\cite{Giuliatti2014} found that the sailboat region in the 
Pluto-Charon system exists only in two intervals: 
$\Delta \omega = \pm 10^\circ$ around $\omega = 0^\circ$ and 
$\Delta \omega = \pm 20^\circ$ around $\omega = 180^\circ$. Our 
results confirm these intervals exist across all mass ratios but 
with varying widths.

\begin{center}
    
\begin{figure}[!h]
    \centering
    \includegraphics[width=0.45\textwidth]{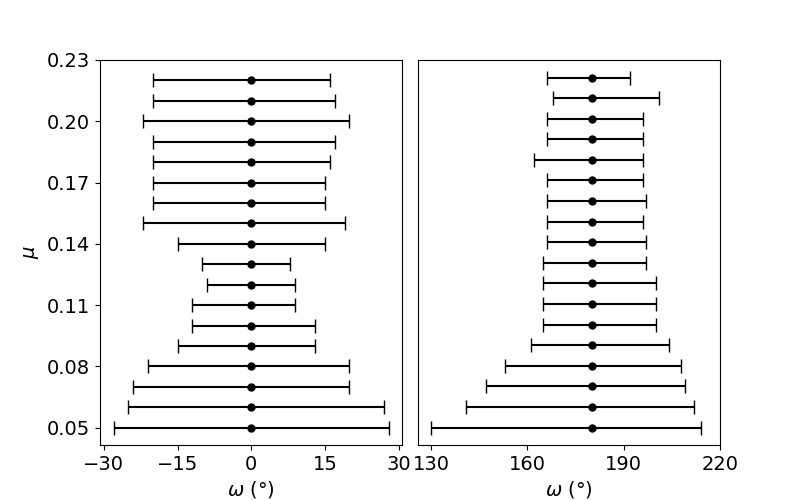}
    \caption{The interval of particle argument of pericenter
    where the sailboat region exists with maximum extension at $\omega = 0^\circ$
    (left plot) and $\omega = 180^\circ$ (right plot).}
    \label{F-Omega}
\end{figure}
\end{center}

Figure~\ref{F-Omega} displays the angular extent of these intervals for 
each mass ratio, with the left panel showing the range around $\omega = 0^\circ$ 
and the right panel around $\omega = 180^\circ$. The intervals exhibit 
asymmetric behavior, differing from the symmetric patterns found in the 
Pluto-Charon system. This asymmetry likely results from our machine 
learning model's generalization across the broader parameter space, 
where the number of stable particles becomes very small at the interval 
boundaries.

The interval widths vary systematically with mass ratio. For 
$\mu < 0.08$, stability exists within $\Delta \omega = \pm 20^\circ$ 
to $\pm 30^\circ$ of $\omega = 0^\circ$. This interval narrows to 
$\pm 10^\circ$ for $\mu = [0.09 - 0.14]$, matching the Pluto-Charon 
findings of \cite{Giuliatti2014}. For $\mu \geq 0.15$, the interval 
widens again to approximately $\pm 20^\circ$.

The behavior around $\omega = 180^\circ$ differs significantly from that 
around $\omega = 0^\circ$. For $\mu \leq 0.09$, the interval ranges from 
$\Delta \omega = \pm 25^\circ$ to $\pm 45^\circ$, while for $\mu \geq 0.09$, 
these intervals narrow to $\Delta \omega \approx \pm 15^\circ$ to $\pm 20^\circ$.

\begin{figure}[!ht]
\centering
    \includegraphics[width=1.\columnwidth]{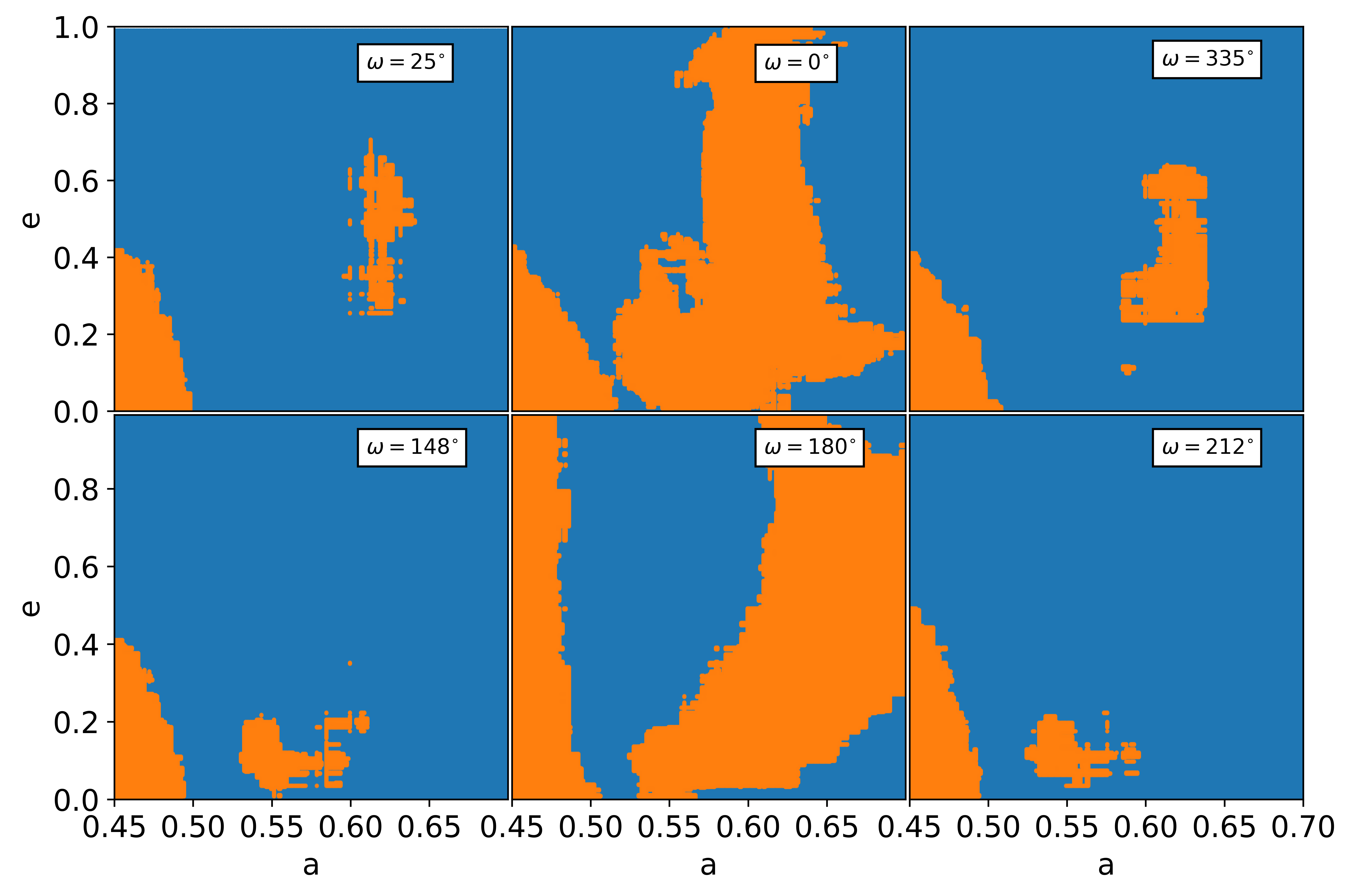}
\caption{Stability maps $(a \times e)$ for a binary system with 
$\mu = 0.05$ showing the sailboat region evolution across different 
values of argument of pericenter. Stable initial conditions are shown 
in orange, unstable in blue.}
\label{F-Seqomega}
\end{figure}

Figure~\ref{F-Seqomega} illustrates the sailboat region evolution for 
$\mu = 0.05$ across six representative values of $\omega$. The maximum 
stable regions occur at $\omega = 0^\circ$ and $180^\circ$, where the 
sailboat exhibits its characteristic shape. At intermediate values 
($\omega = 25^\circ$, $148^\circ$, $212^\circ$, and $335^\circ$), the 
sailboat region either shrinks considerably or disappears entirely, 
confirming the narrow angular windows required for stability.

The geometric constraint underlying this behavior relates to the orientation 
of particle orbits relative to the binary axis. Maximum stability occurs 
when the particle's pericenter aligns with or opposes the primary-secondary 
line, corresponding to $\omega = 0^\circ$ and $180^\circ$. Deviations from 
these orientations disrupt the periodic orbit family that sustains the 
sailboat region, leading to instability for most particles in this 
parameter range.

\section{Stability verification}
\label{S-Poincare}

{
To validate our machine learning predictions and provide independent 
confirmation of the stability patterns identified in previous {sections},  
we employed two complementary verification methods.}

{The stability of the sailboat region was demonstrated by \citet{Giuliatti2014}, 
who studied the sailboat region in the Pluto-Charon system and identified its stability 
through the family BD of periodic orbits described by \citet{Brouke1968}. They employed
Poincaré surfaces of section (PSS) to confirm the region's stability.  
The PSS method is a powerful tool for demonstrating stability by revealing 
the periodic and quasi-periodic trajectories within a system. 
However, its effectiveness is limited by its dependence on the Jacobi Constant 
to reduce the system's degrees of freedom, a constant that only exists in the 
Circular Restricted Three-Body Problem (CRTBP). Additionally, the PSS method's 
representation is constrained to the $x$ and {$v_x$} phase space of a barycentric 
synodic coordinate system, which can limit the scope of analysis.}

{To confirm the stability of the regions obtained by the machine learning 
we present a combination of techniques, including Poincaré surfaces of section (PSS), 
and the Lyapunov Exponent as a chaos indicator. 
The Lyapunov exponent is a fundamental tool for diagnosing chaos in dynamical systems, 
as it quantifies the rate at which nearby trajectories diverge, 
providing a rigorous measure of a system's sensitivity to initial conditions.
% —a key characteristic of chaotic behavior. 
This approach has proven particularly effective for studying chaotic behavior in
celestial mechanics \citep{Popova2013}.  A positive Lyapunov exponent indicates 
the exponential divergence of initially close trajectories, which signifies chaos. 
Conversely, a negative exponent suggests that trajectories converge, indicating 
stability, while a zero value
implies neutral stability. The greater the positive exponent, the more
unpredictable the system becomes over time.}

{In analyzing stochastic regions, it's crucial to estimate the 
time-scale for the onset of chaos in orbital motion, which is defined as the Lyapunov time. 
This is the inverse of the largest Lyapunov Characteristic Number (LCN). 
Obtaining an accurate LCN typically requires a long integration time, on the order of $10^5$
to $10^6$ orbital periods. However, this study uses REBOUND with IAS15 \citep{rebond-ias15}, 
which employs the Mean Exponential Growth factor of Nearby Orbits (MEGNO) method \citep{cincotta2000}. 
MEGNO provides a measure of chaos that is proportional to the LCN, 
allowing for its accurate determination in significantly shorter, more realistic
physical times of approximately $10^3$ periods \citep{cincotta2000}. 
This approach offers a more efficient way to quantify the chaotic nature of these regions.}

{Rather than a comprehensive stability study, we focus on validating 
our machine learning predictions through targeted analysis of specific cases. 
We selected several representative systems for detailed examination using PSS and LCN. 
From Figure \ref{F-Massratio}, which depicts a system where the eccentricity
of the binary is zero and particles lie in plane \(x,y\), 
we chose the cases with $\mu = 0.15$ and $\mu = 0.23$ for analysis using both PSS and LCN.  
For the cases presented in Figure \ref{F-Ecc} ($\mu = 0.12$ and $e_s = 0.023$) 
and Figure \ref{F-InclinationMap} ($\mu = 0.12$ and $i=75^{\circ}$), we used only the LCN
due to the inherent limitations of PSS in analyzing those specific systems. 
It is important to note that these calculations have far greater time and computational 
requirements compared with the machine learning method.}

\begin{figure}[!h]
\centering
  \includegraphics[width=0.35\textwidth]{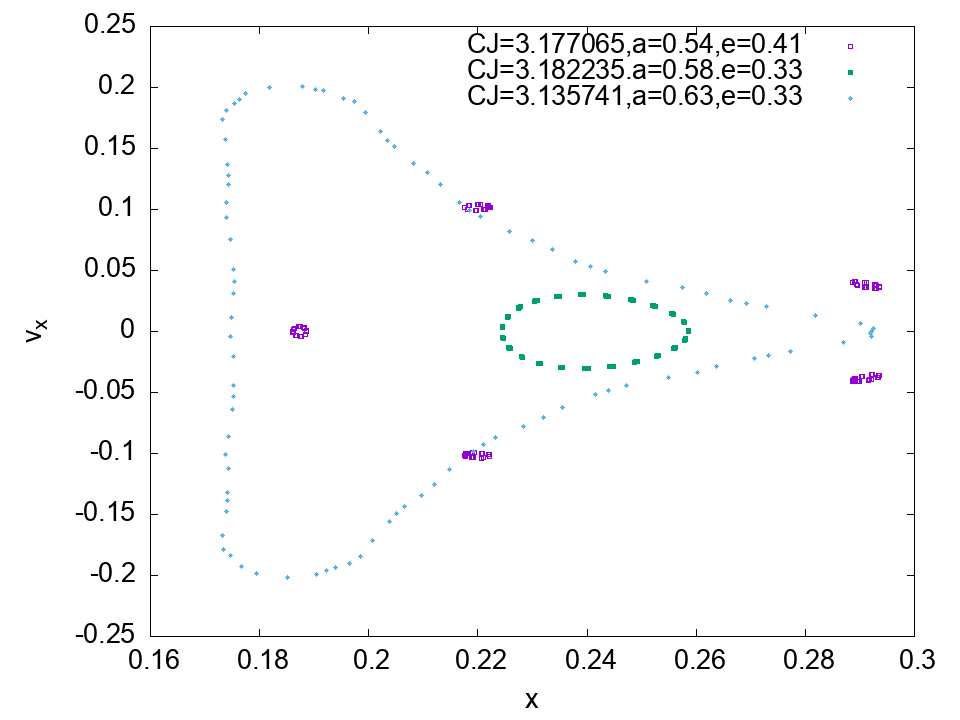} \\
  \includegraphics[width=0.35\textwidth]{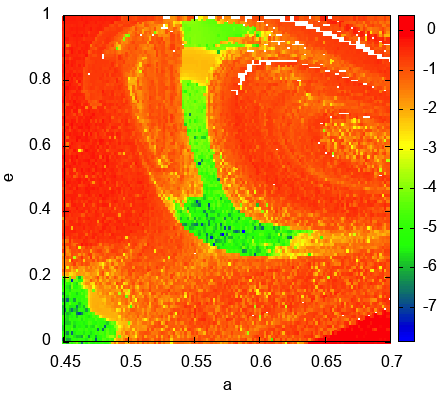}
\caption{Upper panel: Poincaré surfaces of section in the $(x, v_x)$ 
plane for selected trajectories at the boundaries of the sailboat region with 
$\mu = 0.13$. Three trajectories correspond to different positions within the 
sailboat: leftmost edge, center, and rightmost edge. Quasi-periodic orbits 
appear as closed curves, confirming stability. Bottom panel: Map of LCN in the 
same region for case $\mu = 0.13$. The most negative regions indicate less 
chaotic trajectories.}
\label{STAB_C_1}
\end{figure}

{In Figure \ref{STAB_C_1}, we present PSS for trajectories located at 
the boundaries of the sailboat region, as depicted in Figure \ref{F-Massratio}, 
for a mass ratio of $\mu = 0.13$. 
We chose three test particles for projection onto the $(x, v_x)$ plane, 
each having distinct Jacobi constants and positions within the sailboat region: 
one at the far left edge, one at the far right edge, and one in the center. 
The resulting trajectories correspond to quasi-periodic orbits, and their stability 
is confirmed by the regular, predictable patterns observed in the surfaces of
section.  
We also illustrate the same region from Figure \ref{F-Massratio} using the LCN, 
represented by a color gradient. The logarithmic value of the LCN is plotted, 
where more negative values indicate less chaotic trajectories. The similarity between the regions of low 
chaos derived from the LCN and the stability map generated by the machine learning method 
is remarkable, highlighting the efficiency of the latter, 
even when accounting for the initial training time.}

\begin{figure}[!ht]
\centering
    \includegraphics[width=0.35\textwidth]{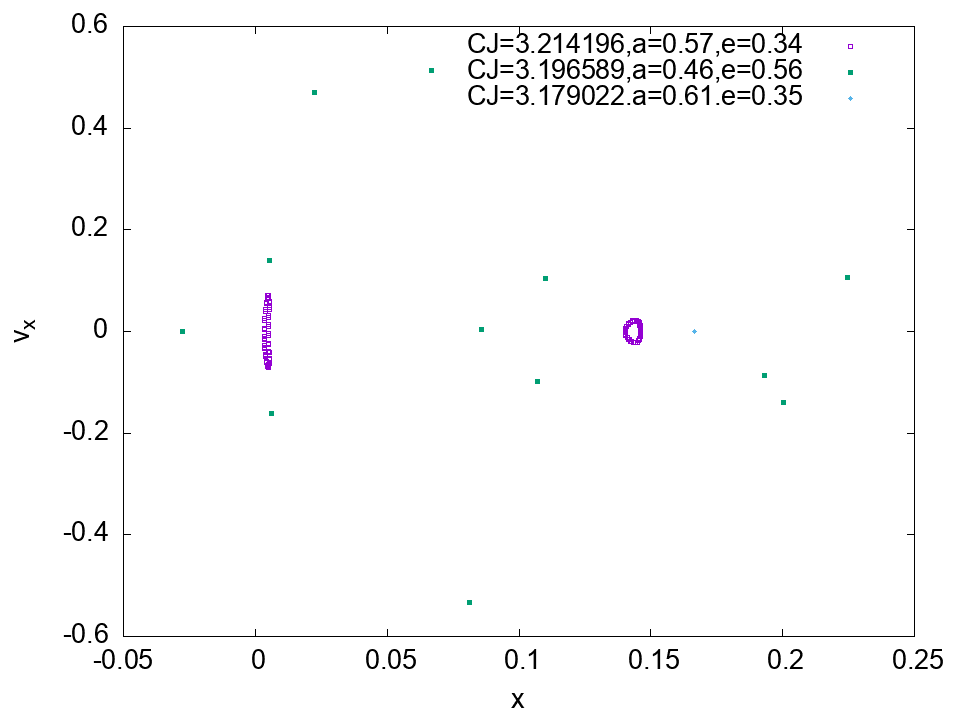}
    \includegraphics[width=0.35\textwidth]{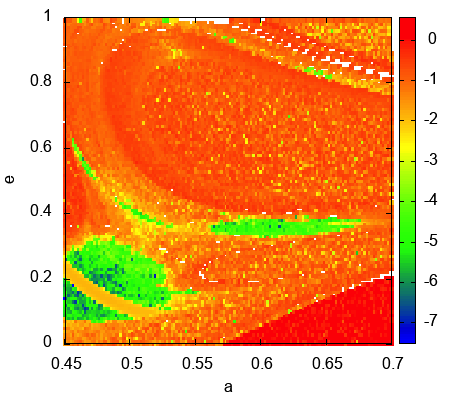}
\caption{Same as Figure \ref{STAB_C_1} for $\mu = 0.23$.}
\label{STAB_C_2}
\end{figure}

{Figure \ref{STAB_C_2} shows an extreme case where $\mu = 0.23$. 
Using PSS, we encountered difficulties in finding quasi-periodic orbits, 
with only one trajectory within the main sailboat region proving stable. 
Despite this challenge with PSS, the LCN method indicates almost the same 
region as more stable compared to our machine learning predictions, with the 
negative logarithm of the LCN being below -4. This implies an LCN below 
$10^{-4}$. This agreement is remarkable given that we trained the machine 
learning method with trajectories that remained stable for $10^4$ orbital 
periods (see section \ref{S-NumericalSimulations}), not with LCN values.}

{When we account for the eccentricity of the binary, we diverge 
from the CRTBP, and the stability of the sailboat regions deteriorates. 
Our machine learning method demonstrates this effect in Figure \ref{F-Ecc}. 
The bottom panel of Figure \ref{STAB_C_2} shows the LCN for the same 
parameter space ($\mu = 0.12$ and $e_s = 0.023$). Comparing these results 
proves more challenging because the LCN regions with more negative values 
are dispersed. However, the stable region remains predominantly located 
in the middle of both figures, indicating correspondence between the 
two methods.}

\begin{figure}[!ht]
\centering
\begin{minipage}{0.48\columnwidth}
    \centering
    \includegraphics[width=\textwidth]{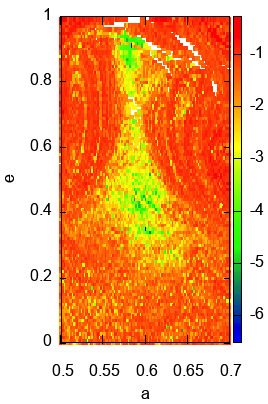}
\end{minipage}%
\hfill%
\begin{minipage}{0.48\columnwidth}
    \centering
    \includegraphics[width=\textwidth]{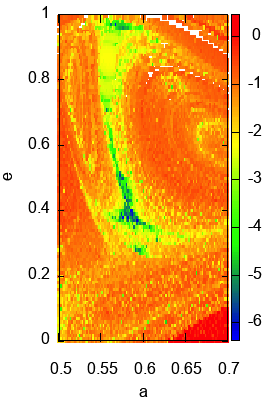}
\end{minipage}
\caption{Lyapunov exponents for regions explored by the ML. Left: region 
of Figure \ref{F-Ecc} ($\mu = 0.12$ and $e_s = 0.023$). Right: region 
of Figure \ref{F-InclinationMap} ($\mu = 0.12$ and $i=75^{\circ}$).}
\label{STAB_C_3}
\end{figure}

{The right panel of Figure \ref{STAB_C_2} shows the region 
corresponding to Figure \ref{F-InclinationMap}, with $\mu = 0.12$ and 
$i = 75^\circ$. \citet{Winter2013} demonstrates that the stability of the 
sailboat region extends out of the plane, even to high inclinations. 
Our machine learning method also identifies this stability, and the LCN 
method shows a region with less chaoticity that is almost compatible. 
Although this comparison shows the most divergence of the four cases 
presented, with our machine learning predictions being more symmetric 
than the LCN results, both methods still reveal similar underlying 
structures. This indicates that our approach can effectively identify 
possible regions of stability, even with certain limitations.}

{These discrepancies reveal that while our machine learning approach 
efficiently maps general stability regions with minimal computational cost, 
some boundary points may be misclassified.}

\section{Final remarks}
\label{S-finalremarks}

The sailboat region is a stable zone for S-type orbits in binary
systems. This region lies approximately midway between the
two massive bodies and supports particles with highly eccentric
orbits reaching $e = 0.9$. {Through stability verification methods 
including Poincar\'e surfaces of section and Lyapunov exponent analysis}, 
we confirmed that this stability originates 
from a family of periodic orbits that persist despite the strong 
gravitational perturbations typical of binary environments.

In this paper, we studied how the sailboat region responds to
changes in binary system parameters. We studied four key parameters: 
binary mass ratio, secondary body eccentricity, particle orbital 
inclination, and argument of pericenter. We combined
direct numerical integration of 1.2 million three-body
systems with machine learning algorithms. This approach allowed us to
predict stability for about $10^9$ initial conditions with accuracy
exceeding 97\%.

Our analysis reveals that the sailboat region exists exclusively within
a limited mass ratio range, $\mu = [0.05, 0.22]$. Below $\mu = 0.05$, the region connects with the
inner stable zone around the primary body. For $\mu > 0.22$,
gravitational perturbations eliminate all
stability in this parameter space.
Within this range, higher $\mu$ values make the sailboat region smaller
and closer to the primary body, while requiring higher minimum
eccentricities.

The secondary body's eccentricity severely constrains sailboat region
existence, in a way that even
 small increases from circular orbits rapidly destabilize
the region, following an exponential decay relationship.
For $\mu = 0.05$, the maximum secondary
eccentricity is about 0.23, but drops to only 0.007 for $\mu = 0.22$.
This strong dependence means that most binary systems with eccentric
components cannot support stable sailboat regions.

Particle inclination shows good tolerance, with the sailboat region
surviving inclinations from $0^\circ$ to $90^\circ$ for most mass ratios.
Retrograde configurations can still support sailboat stability
when $\mu \leq 0.16$, though with reduced spatial
extent. This inclination
tolerance increases the number of possible systems where such regions
might exist.

The argument of pericenter creates strong limits on sailboat region
existence. Stability occurs only within narrow angular ranges:
$\Delta \omega \approx \pm 10^\circ$ to $\pm 30^\circ$ centered
at $\omega = 0^\circ$ and $\omega = 180^\circ$. This selectivity
comes from the periodic orbit family needing specific orientations
relative to the binary axis.

Nevertheless, the method provides a valuable tool for initial 
stability surveys, significantly reducing computational requirements 
for comprehensive dynamical studies. Once trained, our machine learning 
model can reduce computation time up to a factor of $10^{5}$ compared to 
numerical simulations. 
{For instance, it takes about a couple of months using a single 
core with typical specifications to build the training dataset, whereas the model training and ML prediction takes the order of seconds to run. This 
performance allows us }
to explore orbital stability 
across large parameter spaces, though detailed investigations remain 
necessary to confirm the real stability of identified regions.

These results apply to several real systems. Among dwarf planet binaries
in our Solar System, such as Pluto-Charon, Orcus-Vanth and Varda-Ilmar\"e systems have
mass ratios and eccentricities that could support sailboat regions.
These systems are good targets for searching for ring structures or
small satellites in such regions. Our results also help with spacecraft
mission planning in binary environments by identifying stable parking
orbits and regions that need trajectory changes.

\begin{acknowledgements}
We thank the reviewer for their insightful comments and valuable suggestions. 
R.S. acknowledge support by the DFG German Research
Foundation (project 446102036) and 
CNPq (307400/2025-5). 
R.S. and G. R. are grateful to the S\~{a}o Paulo Research Foundation (FAPESP, grant 2024/20150-1). 
T.F.L.L.P. acknowledge support from the Conselho Nacional de Desenvolvimento Científico e Tecnológico (CNPq) - Proc.313994/2025-0.
The authors thank the Improvement Coordination Higher Education Personnel - Brazil (CAPES) - Financing Code 001.
\end{acknowledgements}

\bibliographystyle{aa}
\bibliography{references}

\end{document}